\documentstyle[12pt]{article}

\def\lsim{\;\lower4pt\hbox{$\sim$} \hskip-10pt\raise1.6pt\hbox{$<$}\;}

\def\gsim{\;\lower4pt\hbox{$\sim$} \hskip-10pt\raise1.6pt\hbox{$>$}\;}
\newlength{\abstractwidth}
\begin{document}
\thispagestyle{empty}
\pagestyle{plain}
\renewcommand{\thefootnote}{\fnsymbol{footnote}}
\renewcommand{\thanks}[1]{\footnote{#1}} % Use this for footnotes
\newcommand{\starttext}{
\setcounter{footnote}{0}
\renewcommand{\thefootnote}{\arabic{footnote}}}
%%%%%%%%%%%%%%%%%%%%%%%%%%%%%%%%%%%%%%%%%%%%%%%%%%%%%%%%%%%%%%%

\begin{titlepage}
\rightline{FERMILAB-PUB-93/262-T}
\rightline{Sept. 1993}
\bigskip\bigskip\bigskip

\centerline{\Large\bf Planck-Scale Unification}
\medskip
\centerline{\Large\bf and}
\medskip
\centerline{\Large\bf Dynamical Symmetry Breaking}

\bigskip\bigskip

\centerline{\large\bf Joseph D. Lykken}
\medskip
\centerline{Institute for Theoretical Physics}
\centerline{University of California}
\centerline{Santa Barbara, CA\ \ 93106}

\medskip
\centerline{and}

\medskip

\centerline{Fermi National Accelerator Laboratory}
\centerline{P.O. Box 500}
\centerline{Batavia, IL\ \ 60510}
\bigskip\bigskip

\centerline{\large\bf Scott Willenbrock\thanks{
Present address: Department of Physics, University of Illinois,
1110 West Green St., Urbana, IL 61801}
}
\medskip
\centerline{Fermi National Accelerator Laboratory}
\centerline{P.O. Box 500}
\centerline{Batavia, IL\ \ 60510}

\medskip
\centerline{and}

\medskip

\centerline{Physics Department}
\centerline{Brookhaven National Laboratory}
\centerline{Upton, NY\ \ 11973}

\begin{abstract}
We explore the possibility of unification of gauge couplings near the Planck
scale in models of extended technicolor. We observe that
models of the form $G\times{}$SU(3)$_{c}\times{}$SU(2)$_L\times{}$U(1)$_Y$
cannot be realized, due
to the presence of massless neutral Goldstone bosons (axions) and light
charged pseudo-Goldstone bosons; thus, unification of the known forces near
the Planck scale cannot be achieved. The next
simplest possibility, $G\times{}$SU(4)$_{PS}\times{}$SU(2)$_L\times{}$
U(1)$_{T_{3R}}$, cannot lead
to unification of the Pati-Salam and weak gauge groups near the Planck scale.
However, superstring theory provides relations
between couplings at the Planck scale without the need for an underlying
grand-unified gauge group, which allows unification of the SU(4)$_{PS}$ and
SU(2)$_L$ couplings.
\end{abstract}
\end{titlepage}
\starttext

The standard model of the strong and electroweak interactions is based
on the gauge group ${\rm SU(3)}_c\times {\rm SU(2)}_L\times {\rm
U(1)}_Y$, with ${\rm SU(2)}_L\times {\rm U(1)}_Y$ spontaneously broken
to ${\rm U(1)}_{EM}$ at the weak scale, $(\sqrt2 G_F)^{-1/2}=246$
GeV\null. Although the coupling strengths of the three gauge forces are
apparently unrelated at ordinary energies, it is attractive to
hypothesize that, as a result of their evolution, they are related at
some higher energy \cite{GQW}.  One realization of this conjecture is
grand unification, in which the standard gauge group is embedded in a
larger gauge group, which is spontaneously broken at one or more scales
above the weak scale \cite{GG}. The simplest example is minimal SU(5)
\cite{GG}, which nearly succeeds in unifying the known gauge forces at a scale
of around $10^{15}$ GeV \cite{GQW}, far above the weak scale.

A well-known difficulty with attempts at grand unification is the
enormous disparity between the weak scale and the grand-unified scale.
It is not natural for such a hierarchy of scales to occur if the gauge
symmetries are broken by the vacuum-expectation values of fundamental
scalar fields \cite{GQW,S}. Furthermore, a hierarchy based on
fundamental scalar fields is unstable due to quadratic divergences in
the renormalization of the parameters of the scalar-field potential
\cite{S}. A generic means to stabilize this hierarchy is to invoke
low-energy supersymmetry (SUSY) \cite{SUSY}. Supersymmetry itself must
be softly broken, but at a scale not far above the weak scale if it is to
protect the hierarchy.

The introduction of supersymmetry requires the existence of the
superpartners of the standard particles, with masses of order the SUSY
breaking scale, as well as an additional Higgs doublet and its superpartner.
These additional particles influence the evolution of
the three gauge couplings \cite{DRW}. As is well known, minimal SUSY
SU(5) succeeds in unifying the three known gauge forces, at a scale of about
$10^{16}$ GeV \cite{GUT}. This is often considered to be indirect evidence of
the fundamental correctness of both SU(5) grand unification and supersymmetry.

The other known force, gravity, is not a gauge interaction.
At ordinary energies, gravity is described by a classical field theory.
The scale at which quantum gravity becomes relevant is $(8\pi
G_N)^{-1/2} \approx 2.4\times10^{18}$ GeV, which we will refer to as the
Planck scale.\footnote{The energy $G^{-1/2}_N=1.22\times 10^{19}$ GeV is
usually called the Planck scale. The factor $8\pi$ comes from the Einstein
field equation, $G^{\mu\nu}=8\pi G_NT^{\mu\nu}$.}
It is compelling to hypothesize that
this is a fundamental scale of physics, and that unification of the four
known forces should occur there. The fact that the minimal SU(5)
grand-unified scale is close to the Planck scale also suggests that
gravity and unification are related \cite{GQW}.

Despite the success of the minimal SUSY SU(5) grand-unified scenario, we
wish to explore models of Planck-scale unification based on dynamical
symmetry breaking \cite{S,W,FS}. There are several motivations for
doing so. First, dynamical symmetry breaking is the only other known
generic mechanism besides supersymmetry to maintain the hierarchy
between the Planck scale (or grand-unified scale) and the weak scale
\cite{GQW,S}. Thus it is the only realistic alternative to the
SUSY grand-unified scenario. Second, it {\it explains\/} why these
scales are so enormously different \cite{S}.  Third,
the SU(3)$_c$ and SU(2)$_L$ couplings merge at
about $10^{17}$ GeV in the standard model, close to the Planck scale,
if the Higgs doublet is
removed from the evolution equations.\footnote{This is with two-loop evolution
and the strong coupling $\alpha_3(M_Z)=.115$.}
This suggests replacing the Higgs
sector with some other electroweak-symmetry-breaking mechanism.
Fourth, superstring theory
predicts relations between couplings at the Planck scale {\it without
the need for an underlying grand-unified gauge group\/} \cite{GI}. This
opens up the possibility of Planck-scale unification with dynamical
symmetry breaking, which may be impossible in a grand-unified approach
\cite{FS}\cite{FS2}-\cite{GR}.

Since we are attempting to relate physics at the weak scale to physics
at the Planck scale, we must consider models of dynamical symmetry
breaking that account for the generation of
fermion masses as well as the weak-boson masses.
One such class of models is extended technicolor (ETC)
\cite{EL,DS}.\footnote{For a discussion of gauge- and Yukawa-coupling
unification in a SUSY top-quark-condensate model, see Ref.~\cite{WAG}.}
These models
have several well-known potential problems: large flavor-changing
neutral currents \cite{EL,DRK,DE}, large contributions to low-energy
precision electroweak phenomena \cite{PT}, and relatively light
pseudo-Goldstone bosons \cite{EL,DS}. We will not address these
problems, but simply assume they may be obviated via fixed-point or
walking technicolor
\cite{WT}, or some other mechanism.  The lack of any realistic model
is another difficulty with extended technicolor.

There is one potential problem with extended-technicolor models which
cannot be ignored: the presence of massless neutral Goldstone bosons
(weak-scale axions \cite{WW}) and light charged pseudo-Goldstone
bosons, of mass ${\cal O}(\alpha M_Z)\sim 5$ GeV \cite{EL,D}. The
necessary and sufficient conditions on the ETC representation for the
avoidance of these particles  were derived long ago by Eichten and Lane
\cite{EL}. They showed that there may be at most one irreducible
representation (irrep) of SU(2)$_L$ doublets, ${\cal D}_L$, and at most
two (inequivalent) irreps of SU(2)$_L$ singlets, ${\cal D}_{u^c}$ and
${\cal D}_{d^c}$, with SU(2)$_L$ singlet leptons belonging to one or
both of these.\footnote{In Ref.~\cite{EL}, ${\cal D}_L$, ${\cal
D}_{u^c}$, and ${\cal D}_{d^c}$ are called ${\cal D}^S_L$, ${\cal
D}^S_{u_R}$, and ${\cal D}^S_{d_R}$, respectively ($S=$``sideways"). We
have chosen to work with left-handed fermions.} SU(2)$_L$ may or may not
commute with the extended-technicolor group.

Using these conditions, it is easy to enumerate the grand-unified
models based entirely on dynamical symmetry breaking which are
potentially realistic.  There can be at most one irrep of the (simple)
grand-unified gauge group, since more than one irrep would produce an
ETC representation which violates the above conditions \cite{EL}. This
irrep must be complex to avoid unification-scale masses \cite{G}. In
order for the grand-unified group to break itself via tumbling
\cite{RDS}, the coupling must become strong as one descends from the
unification scale, so the theory must be asymptotically free. The only
anomaly-free, irreducible, complex representations of simple groups
which are also asymptotically free are the 16, 126, and 144-dimensional
representations of SO(10); the 64-dimensional representation of SO(14);
the 256-dimensional representation of SO(18); and the 27-dimensional
representation of $E_6$ \cite{EKK}. The group SO(10), of rank 5, is not
large enough to accommodate the standard gauge group, of rank 4, and a
technicolor group. The 27-dimensional representation of $E_6$ can
accommodate only one generation of fermions. The 64-dimensional
representation of SO(14) can accommodate only four generations,
which is not enough to support a non-Abelian technicolor
group\footnote{For a two-generation model based on SO(14), with SU(2)
technicolor, see Ref.~\cite{FS}. See the second note added to that
paper.}. This leaves the 256-dimensional spinor representation of
SO(18). A grand-unified technicolor model based on this group and
representation has been considered in Refs.~\cite{GRS,WZ}, and more
recently in Ref.~\cite{GR}; see also Ref.~\cite{AT}.  The group
SO(10)${}\times{}$SO(10), with a discrete symmetry equating the
couplings and the representation (16,16),
is also a candidate since as
many as 22 16-dimensional representations are allowed by asymptotic
freedom \cite{EKK}. A model based on this
group and representation has been considered in Ref.~\cite{DMW}.  A model
based on this group and the reducible representation $(16,10)\oplus(10,16)$,
which is asymptotically free,
has been considered in Refs.~\cite{DMW2,KING2}; h
owever, it suffers from light color-singlet Goldstone bosons.

One need not insist that the breaking of the grand-unified gauge group
be dynamical. As long as this breaking occurs near the Planck scale, it
may be produced by the vacuum-expectation value of a fundamental scalar
field without requiring an unnatural hierarchy of scales.\footnote{However,
the small observed value of the cosmological constant remains a mystery.}
It is only the
breaking of the electroweak interaction which must
proceed dynamically in order to produce and stabilize a hierarchy of
scales \cite{GQW,S}. Thus we need not insist that the irrep  of the
grand-unified group be asymptotically free. Nevertheless, the
restriction to an anomaly-free, irreducible, complex representation of
the grand-unified gauge group is a severe constraint. Only complex
representations of $E_6$ and spinor
representations of SO($4N+2$) [$N\ge2$] are generically
allowed. For SU($N$), the lowest-dimensional anomaly-free, irreducible,
complex representation is the 374,556-dimensional representation of
SU(6) \cite{EKK}.

Rather than pursuing grand-unified technicolor models from the top down
any further, we will instead consider such models from the bottom up.
Another consequence of the representation content of extended
technicolor models is that quarks and leptons cannot reside in separate
representations. This implies that SU(3)$_c$ and U(1)$_Y$ cannot survive
as ununified groups above the ETC scale \cite{EL}. Thus, {\it the
observed fact that ${\rm SU(3)}_c\times {\rm SU(2)}_L \times {\rm
U(1)}_Y$ (nearly) unify at around $10^{15}$ GeV is an accident if nature is
described by an extended-technicolor model}. Put another way,
in extended-technicolor theories one necessarily
loses the successful prediction of the weak mixing angle \cite{GQW,GUT}.
In searching for
Planck-scale unification of the low-energy forces, one must therefore consider
groups which contain SU(3)$_c$ and U(1)$_Y$ as subgroups.\footnote{In
Ref.~\protect\cite{F}, a class of grand-unified technicolor models of the
form SU($N$)$\to$SU($n$)$_{TC}{}\times{}$SU(3)$_c\times{}$SU(2)$_L
\times{}$U(1)$_Y$ are ruled out based on anomaly cancellation and
asymptotic freedom. Such models do not provide fermion masses, so we do not
consider them.}
The simplest manner to achieve this, and one that is often employed in
model building \cite{DRS,DRK}\cite{BCS}-\cite{AW}\cite{DMW}-\cite{KING},
is to embed SU(3)$_c\times{}$U(1)$_Y$
in a Pati-Salam group \cite{PS}, SU(4)$_{PS}\times{}$U(1)$_{T_{3R}}$,
where the U(1)$_{T_{3R}}$ quantum numbers are chosen such that the
standard particles have the correct hypercharge when
SU(4)$_{PS}\times{}$U(1)$_{T_{3R}}$ is
broken. Alternatively, U(1)$_{T_{3R}}$ may be the diagonal subgroup of
an SU(2)$_R$ group.  Quarks and leptons reside in the four-dimensional
representation of SU(4)$_{PS}$, with the leptons providing the fourth
``color'' \cite{PS}. We will pursue models of the form
$G\times{}$SU(4)$_{PS}\times{}$SU(2)$_L\times{}$U(1)$_{T_{3R}}$, where
$G$ contains the ETC group, and attempt to unify the Pati-Salam and weak
couplings near the Planck scale.

The bound $BR(K_L\to\mu e) <3.3\times10^{-11}$ (Ref.~\cite{E791})
implies that $M_{PS}/g_{PS}$ $\gsim 10^6$ GeV (Ref.~\cite{DRK,DJ}). The
contribution of the broken Pati-Salam generators to the mass of the
axion and the charged pseudo-Goldstone boson is therefore${}\lsim1$ GeV
(Ref.~\cite{DRK}). This may be increased in walking technicolor by as
much as $M_{ETC}/\Lambda_{TC}$ \cite{WT}. Assuming
$M_{ETC}/\Lambda_{TC}\lsim 10^3$, the allowed range of $M_{PS}$ is therefore
about $10^6$--$10^7$ GeV\null.

The model we study has the representation content ($G$, SU(4)$_{PS}$,
SU(2)$_L$, U(1)$_{T_{3R}}$) of \cite{DRS}

\begin{eqnarray}
{\cal D}_L & = & (N_g,4,2,0) \nonumber \\
{\cal D}_{u^c} & = & (N_g, \bar4, 1, -\frac{1}{2}) \nonumber \\ {\cal
D}_{d^c} & = & (N_g, \bar4, 1, +\frac{1}{2}) \label{REP} \end{eqnarray}

\noindent We leave $G$ unspecified, since we only need the dimension of
the representations, i.e., the number of generations of fermions and
technifermions, $N_g$. $G$ need not be simple, and may contain groups other
than extended technicolor. This is the unique representation which is free
of SU(4)$_{PS}\times{}$U(1)$_{T_{3R}}$ anomalies and contains no exotic
representations.
$G$ anomalies may be canceled by adding representations which are
SU(4)$_{PS}\times{}$SU(2)$_L\times{}$U(1)$_{T_{3R}}$ singlets, if
needed.  Such representations may also be needed to break the
extended-technicolor group dynamically \cite{DRS}.

The one-loop renormalization-group evolution equation for the couplings
is \begin{equation} \frac{1}{\alpha_n(\mu)} - \frac{1}{\alpha_n(\mu_0)}
= - \frac{b_n}{2\pi}\; \ln \frac{\mu}{\mu_0} \label{RNG} \end{equation}

\noindent where $\alpha_n=g^2_n/4\pi$, and $b_n$ is the one-loop
beta-function coefficient,

\begin{equation}
b_n=  - \frac{11}{3} C_2(G) + \frac{2}{3} \sum_R T(R) \label{BETA}
\end{equation}

\noindent where $C_2(G)$ is the quadratic Casimir of the group, and
$T(R)$ is the Dynkin index of the (chiral) representation $R$.
We equate the Pati-Salam and weak couplings at the
unification scale, $M_U$, and evolve the couplings down to the
Pati-Salam scale, $M_{PS}$, using the beta-function coefficients

\begin{eqnarray}
b_4 &=& - \frac{44}{3} + \frac{4}{3} N_g \label{B4} \\ b_2 &=&
- \frac{22}{3} + \frac{4}{3} N_g \;.\label{B2} \end{eqnarray}

\noindent At $M_{PS}$, SU(4)$_{PS}\times{}$U(1)$_{T_{3R}}$ breaks down to
SU(3)$_c\times{}$U(1)$_Y$. The strong coupling, $\alpha_3$, equals the
Pati-Salam coupling, $\alpha_4$, at this scale and evolves down to the
weak scale with the beta-function coefficient

\begin{equation}
b_3=-11+\frac{4}{3}N_g \;.\label{B3}
\end{equation}

\noindent At the scale
$\Lambda_{TC}$ the technicolor force becomes strong and breaks
SU(2)$_L\times{}$U(1)$_Y$ to U(1)$_{EM}$. Scaling from QCD and SU($N$)
technicolor in the large $N$ limit, one finds \cite{FS}

\[
\Lambda_{TC} = \frac{(\sqrt2 G_F)^{-1/2}}{f_\pi}\; \Lambda_{QCD} \left(
\frac{3}{N}\right)^{1/2} \; \frac{1}{r^{1/2}} \approx (520\mbox{ GeV})
\left( \frac{3}{N}\right)^{1/2}\; \frac{1}{r^{1/2}}  \]

\noindent for $r$ technidoublets. For one technigeneration $(r=4)$
and $N\ge 2$ one finds $\Lambda_{TC}\le 300$ GeV\null. Technifermions
acquire a dynamical mass of this order, and decouple from the
renormalization-group evolution below this scale. Pseudo-Goldstone
bosons lighter than $\Lambda_{TC}$ do contribute to the beta-function
coefficients, but the uncertainty in their masses does not permit us to
include them. Since $\Lambda_{TC}$ is not far above $M_Z$, where the
couplings are known, neglecting the contributions of the pseudo-Goldstone
bosons introduces only a small error.\footnote{We have verified this by
including the pseudo-Goldstone bosons of a one-technigeneration
model, with the masses estimated in Ref.~\cite{FS}.}
Thus, below $\Lambda_{TC}$ we evolve the couplings down to $M_Z$ with
the beta-function coefficients, $b^{SM}_n$, of the three known
generations of quarks and leptons.

Putting it all together yields a
relation between the couplings at $M_Z$:\footnote{We are neglecting the
fact that $m_t>M_Z$. For $m_t<200$ GeV, this introduces only a small
error.}

\begin{equation}
\frac{1}{\alpha_2(M_Z)} - \frac{1}{\alpha_3(M_Z)} = \frac{11}{6\pi}
\left[ 2\ln \frac{M_U}{\Lambda_{TC}} - \ln \frac{M_{PS}}{\Lambda_{TC}} +
\ln \frac{\Lambda_{TC}}{M_Z} \right] \;.\label{SIMPLE} \end{equation}

\noindent Note that $N_g$ has canceled out; the fermions do not contribute to
the relative evolution of $\alpha_2$ and $\alpha_3$, nor $\alpha_2$ and
$\alpha_4$.  Using

\begin{eqnarray}
\alpha_3(M_Z) &=& .115\pm.010 \nonumber \\
\alpha_2(M_Z) &=& \frac{\alpha(M_Z)}{\sin^2\theta_W(M_Z)} =
\frac{1}{29.7} \label{INITIAL}
\end{eqnarray}

\noindent it is easy to show that Eq.~(\ref{SIMPLE}) cannot be satisfied
for any value of $M_{PS}$ between $10^6$--$10^7$ GeV and $M_U$ between
$10^{14}$--$10^{18}$ GeV\null. Thus
{\it SU(4)$_{PS}$ and SU(2)$_L$ cannot be unified into a larger group
near the Planck scale.}  The reason for this observation is
simple. In the standard model
with no Higgs doublet, the SU(3)$_c$ and SU(2)$_L$ couplings
meet at about $10^{17}$ GeV, not far from the Planck scale. When SU(3)$_c$
is subsumed by SU(4)$_{PS}$ at $M_{PS}$, the beta-function coefficient
decreases by $-11/3$, driving the Pati-Salam coupling much lower than the
SU(2)$_L$ coupling near the Planck scale.

Faced with the failure to grand-unify the Pati-Salam and weak gauge groups
near the Planck scale, we turn to string unification of gauge couplings.
Superstring theory is the leading candidate for a quantum theory of
gravity. Although supersymmetry is necessary for a consistent string
theory, it need not survive to low energies, and may be broken at the
Planck scale.\footnote{However, the fact that the cosmological constant
vanishes in an exactly supersymmetric theory can be used to argue that
SUSY should survive to low energies \cite{DRSW}.}
A generic feature of superstring theory
is tree-level relations between couplings at the string-unification scale,
without the need for a grand-unified gauge group.\footnote{For a review, see
Refs.~\cite{I,DALLAS}.} These relations follow
from the need to embed the gauge symmetry into a unitary,
modular-invariant conformal field theory \cite{GI}. The relations are of
the form

\begin{equation}
k_ng^2_n = g^2_{\rm string} \label{LEVEL} \end{equation}

\noindent where $k_n$ is the level of the Kac-Moody algebra associated
with the gauge group with coupling $g_n$ at the string-unification
scale, and $g_{\rm string}$ is the string coupling. The levels are positive
integers for non-Abelian groups. The higher the level, the larger the
allowed representations of the gauge group (e.g., for SU($N$)
the Dynkin labels of the
representations must sum to less than or equal to $k_n$). The levels for
Abelian groups may take any rational value. String
unification not only
allows a more liberal condition for relating couplings near the Planck
scale, it also frees one from the constraint that the fermions must form
a single irrep of the grand-unified gauge group in extended technicolor.
Even if superstring theory should ultimately
prove not to be realized in nature, it provides an existence proof of
Planck-scale unification other than grand unification.

The string scale, $M_{\rm string}$, is related
to the Planck scale, $M_P=(8\pi G_N)^{-1/2}$, by

\begin{equation}
M_{\rm string} = g_{\rm string} M_P \label{MSTRING}
\end{equation}

\noindent at tree level. An estimate of the effect of Planck-scale
physics (threshold effect) on the scale $M_U$ at which the couplings most
closely satisfy Eq.~(\ref{LEVEL}) is \cite{K}

\begin{equation}
M_U = \frac{e^{(1-\gamma)/2} 3^{-3/4}}{\sqrt{2\pi}} M_{\rm string} \approx
.2 \; M_{\rm string}\;. \label{MUNI}
\end{equation}

\noindent Due to the uncertainty in this estimate, we will
vary the unification scale $M_U$ between
$10^{17}$--$10^{18}$ GeV\null. The fact that the minimal SUSY SU(5)
grand-unification scale is about $10^{16}$ GeV may be construed as a
deficiency of the model from the perspective of string theory
\cite{I2,AEKN,BL,DALLAS}.

Relating the SU(4)$_{PS}$ and SU(2)$_L$ couplings at the unification scale
via Eq.~(\ref{LEVEL})
and evolving the couplings as before yields the relation

\begin{eqnarray}
\frac{1}{k_4\alpha_3(M_Z)} - \frac{1}{k_2\alpha_2(M_Z)} & =
& \frac{1}{2\pi} \Biggl[ \left( \frac{b_4}{k_4} - \frac{b_2}{k_2}
\right) \ln \frac{M_U}{\Lambda_{TC}} - \frac{(b_4-b_3)}{k_4} \ln
\frac{M_{PS}}{\Lambda_{TC}} \Biggr. \nonumber\\ & & \quad \Biggl. -
\left(\frac{b^{SM}_2}{k_2} - \frac{b^{SM}_3}{k_4} \right) \ln
\frac{\Lambda_{TC}}{M_Z} \Biggr]\;. \label{MASTER} \end{eqnarray}

\noindent For $k_2=k_4=1$, Eq.~(\ref{MASTER}) reduces to Eq.~(\ref{SIMPLE}).
Thus {\it unification of the Pati-Salam and weak couplings cannot be
achieved with unit Kac-Moody levels.}  From Eq.~(\ref{MASTER}) we see
that this statement is true for $k_2=k_4$ in general.

It is possible to construct string models
with different groups realized at different levels \cite{L,FIQ}.
Equation~(\ref{MASTER}) may be solved for
$k_4/k_2$, varying $M_{PS}$ between $10^6$--$10^7$ GeV and $M_U$
between $10^{17}$--$10^{18}$ GeV\null. The variation of $\alpha_3$
within the range of Eq.~(\ref{INITIAL}) is a small effect. We find the
values of $k_4/k_2$ given in Table~1 for various choices of $N_g$.
Only $N_g=8$ yields a model (nearly) consistent with $k_4=2$, $k_2=1$.
If SU(4)$_{PS}$
and SU(2)$_L$ are realized at different levels, they cannot be subgroups of
the same group (such as SO(10)).  For $N_g\ge 10$, the SU(2)$_L$ coupling
blows up before $10^{17}$ GeV.  The Pati-Salam coupling is asymptotically
free for $N_g\le 10$.

We may also evolve the U(1)$_{T_{3R}}$ coupling,
$\alpha_{1R}$, up to the Planck
scale and find its relation to $\alpha_2$ and $\alpha_4$.  The hypercharge
generator is related to the U(1)$_{T_{3R}}$ generator by\footnote{The
hypercharge generator is normalized such that $Q=T_{3L}+Y$.}
\begin{equation}
Y=T_{3R}+\sqrt{\frac{2}{3}}P_{15}
\label{HYPER} \end{equation}
where $P_{15}$ is the SU(4)$_{PS}$ generator $P_{15}=1/\sqrt{24}\;
{\rm diag}(1,1,1,-3)$.
The coupling $\alpha_{1R}$ is related to the hypercharge coupling, $\alpha_1$,
at $M_{PS}$ by
\begin{equation}
\alpha_1=\frac{\alpha_{1R}\alpha_3}
{\alpha_3+\frac{2}{3}\alpha_{1R}}\;.
\label{ALPHA1} \end{equation}
Evolving the couplings as above, and using $\alpha_1(M_Z)=\alpha(M_Z)/\cos^2
\theta_W=1/98.2$, yields the values of $k_4/k_{1R}$ given in Table 1.
The value $k_4/k_{1R}=1$ would suggest that U(1)$_{T_{3R}}$ and SU(4)$_{PS}$
are subgroups of SO(10), broken at $M_U$; this value is (nearly) obtained
for $N_g\le 9$ (the lower end of the range corresponds to $M_{PS}=10^6$ GeV,
$M_U=10^{17}$ GeV).  In a specific model, one could also evolve the ETC
coupling up to the unification scale and see if it has a simple relation to
the other couplings.

\bigskip

\begin{center}
\begin{tabular}{cccc}
\multispan3 \hfill Table 1 \hfill\\ \\ $N_g$ &
$k_4/k_2$ & $k_4/k_{1R}$  \\
5 & 1.37--1.49 & 1.06--1.23 \\
6 & 1.50--1.65 & 1.07--1.28 \\   7 & 1.65--1.89 & 1.08--1.37  \\
8 & 2.04--2.59 & 1.11--1.54 \\   9 & 3.58--5.48 & 1.16--2.00
\\  \end{tabular}   \end{center}

Although $k_4/k_2$ may take any rational value in principle, the fact that the
fermions lie in the fundamental representations of the gauge groups suggests
that the levels are small.  Furthermore, in specific models the levels
are restricted by other considerations\cite{FIQ}, such as
the fact that the central charges of the Kac-Moody factors must sum
to $\le 22$.
For example, consider $N_g=8$ with SU(8) extended technicolor, and with
SU(4)$_{PS}\times{}$U(1)$_{T_{3R}}$ as subgroups of SO(10), broken at $M_U$.
The central charge of a level $k_n$ Kac-Moody algebra of the group $G$ is
\begin{equation}
c_n = \frac {k_n {\rm dim}(G)}{k_n+C_2(G)}\;.
\label{CENTRAL} \end{equation}
For SU(8) realized at level 1, $c_8=7$. For $SU(2)_L$ realized at level 1,
$c_2=1$. Thus $c_{10}$ must be $\le 14$,
which implies $k_{10}\le 3$.

The above analysis is accurate to one-loop order.  Attempts to refine it must
deal with several issues besides the extension of the beta functions to two
loops.  We have already mentioned the pseudo-Goldstone-boson
contribution to the beta-function coefficients.  The proper treatment of the
threshold due to the dynamical technifermion mass is more complicated than the
simple step function we used.  The technicolor force influences the evolution
of the other couplings at two loops, and may have a significant effect,
especially if it ``walks'', i.e., remains strong over an order of magnitude
or more in energy.

In extended technicolor, one has in mind that there are several
symmetry-breaking scales above the weak scale, and that these are ultimately
responsible for the hierarchy of the masses of the three known generations of
fermions.  However, it is not implausible that the weak force remains
ununified up to the Planck scale.  We have seen that this cannot be the case
for SU(3)$_c\times{}$U(1)$_Y$; however, it is possible for
SU(4)$_{PS}\times{}$U(1)$_{T_{3R}}$.

It is striking that the known fermions form representations of the group
SU(5) (and also SO(10)); this alone is compelling support for SU(5)
(and perhaps even SO(10)) grand unification.  Since SU(5) is eschewed
in our string-unified model (and also SO(10), from the perspective
of SU(4)$_{PS}$ and SU(2)$_L$ unification), this may be regarded as
a deficiency of this approach.  However, the hypercharge quantum numbers
of the known fermions may be fixed by the requirement of anomaly
cancellation alone (including the mixed gravitational anomaly), without
recourse to grand unification \cite{ANOM}.  Perhaps the quantum
numbers of the known fermions reflect something other than SU(5) or
SO(10) grand unification.

As we remarked in the introduction, the SU(3)$_c$ and SU(2)$_L$ couplings
merge at about $10^{17}$ GeV, close to the Planck scale, if the Higgs
doublet is removed from the standard model.  Our attempt to implement
this by replacing the Higgs doublet with a generation of technifermions
was foiled by the need to break the chiral flavor symmetry in order to
generate fermion masses.  In the minimal SUSY
SU(5) grand-unified model, it is actually the addition of a second Higgs
doublet and the superpartners of both Higgs doublets which are responsible
for the unification of the couplings; the superpartners
of the other particles (in particular, the gauginos) merely increase the
unification scale \cite{DRW}.
This again suggests that it is the electroweak-symmetry-breaking sector
which is responsible for producing a successful unification of the
couplings.  We will never be confident of our extrapolations up to the
Planck scale until we understand the electroweak- and
flavor-symmetry-breaking mechanisms.

In this letter we have remarked that
SU(3)$_c$ and U(1)$_Y$ cannot survive as ununified groups up to the Planck
scale in extended-technicolor models, so the observed (near) unification
of ${\rm SU(3)}_c\times {\rm SU(2)}_L\times {\rm U(1)}_Y$ near the
Planck scale in minimal SU(5) cannot be realized in these models.
The simplest models, based on embedding SU(3)$_c\times{}$U(1)$_Y$ into
SU(4)$_{PS}\times{}$U(1)$_{T_{3R}}$, cannot
unify the Pati-Salam and weak gauge groups near the Planck scale.
However, superstring theory provides relations
between couplings at the Planck scale without the need for an underlying
grand-unified gauge group, which allows unification of the SU(4)$_{PS}$ and
SU(2)$_L$ couplings.
\bigskip

\leftline{\bf Acknowledgements}
\medskip

We are grateful for conversations with  T.~Appelquist, S.~Chaudhuri,
E.~Eichten, B.~Holdom, V.~Kaplunovsky, K.~Lane, W.~Marciano, S.~Raby,
and J.~Terning. The work of J.L. was supported in part by the NSF under
Grant No. PHY89-04035. S.W.\ thanks the Aspen Center for Physics where part
of this work was performed. S.W.\ was supported under contract no.\
DE-AC02-76CH00016 with the U.S.\ Department of Energy, and partially
supported by the Texas National Research Laboratory Commission.


\begin{thebibliography}{99}

\bibitem{GQW} H.~Georgi, H.~Quinn, and S.~Weinberg, Phys.\ Rev.\ Lett.\
{\bf 33}, 451 (1974).

\bibitem{GG} H.~Georgi and S.~Glashow, Phys.\ Rev.\ Lett.\ {\bf32}, 438
(1974).

\bibitem{S} L.~Susskind, Phys.\ Rev.\ D {\bf20}, 2619 (1979).

\bibitem{SUSY} M.~Veltman, Acta Phys.\ Polon.\ B {\bf12}, 437 (1981);
L.~Maiani, {\it Proceedings of the Summer School on Particle Physics},
Gif-Sur-Yvette, 1979, (Paris, 1980), p.~1; E.~Witten, Nucl.\ Phys.\
{\bf B188}, 513 (1981).

\bibitem{DRW} S.~Dimopoulos, S.~Raby, and F.~Wilczek, Phys.\ Rev.\ D
{\bf24}, 1681 (1981); M.~Einhorn and D.~R.~T.~Jones, Nucl.\ Phys.\
{\bf B196}, 475 (1982); W.~Marciano and G.~Senjanovic, Phys.\ Rev.\ D
{\bf 25}, 3092 (1982).

\bibitem{GUT} J.~ Ellis, S.~Kelley, and D.~Nanopoulos, Phys.\ Lett.\
{\bf260B}, 131 (1991); U.~Amaldi, W.~de~Boer, and H.~F\"urstenau, Phys.\
Lett.\ {\bf260B}, 447 (1991); P.~Langacker and M.~Luo, Phys.\ Rev.\ D
{\bf44}, 817 (1991).

\bibitem{W} S.~Weinberg, Phys.\ Rev.\ D {\bf13}, 974 (1976); D {\bf19},
1277 (1979).

\bibitem{FS} E.~Fahri and L.~Susskind, Phys.\ Rep.\ {\bf74}, 277 (1981).

\bibitem{GI} P.~Ginsparg, Phys.\ Lett.\ {\bf197B}, 139 (1987).

\bibitem{FS2} E.~Fahri and L.~Susskind, Phys.\ Rev.\ D {\bf20}, 3404
(1979).

\bibitem{F} P.~Frampton, Phys.\ Rev.\ Lett.\ {\bf43}, 1912 (1979).

\bibitem{GRS} M.~Gell-Mann, P.~Ramond, and R.~Slansky, in {\it Supergravity},
eds. P.~van~Nieuwenhuizen and D.~Freedman (North-Holland, Amsterdam, 1979),
p.~315.

\bibitem{WZ} F.~Wilczek and A.~Zee, Phys.\ Rev.\ D {\bf 25}, 553 (1982).

\bibitem{DMW} A.~Davidson, P.~Mannheim, and K.~Wali, Phys.\ Rev.\ Lett.\
{\bf 45}, 1135 (1980).

\bibitem{DMW2} A.~Davidson, P.~Mannheim, and K.~Wali, Phys.\ Rev.\ Lett.\
{\bf 47}, 149 (1981); Phys.\ Rev.\ D {\bf 26}, 1133 (1982); F.~del~Aguila
and A.~Mendez, Z.\ Phys.\ C {\bf 13}, 347 (1982).

\bibitem{KING} S.~King, Phys.\ Lett.\ {\bf 184B}, 49 (1987).

\bibitem{KING2} S.~King, Nucl.\ Phys.\ {\bf B320}, 487 (1989).

\bibitem{AT} T.~Appelquist and J.~Terning, in
{\it Proceedings of the International Workshop on
Electroweak Symmetry Breaking}, Hiroshima, 1991, eds. W.~Bardeen, J.~Kodaira,
and T.~Muta (World Scientific, Singapore, 1992), p.~68.

\bibitem{GR} G.~Giudice and S.~Raby, Nucl.\ Phys.\ {\bf B368}, 221
(1992).

\bibitem{EL} E.~Eichten and K.~Lane, Phys.\ Lett.\ {\bf90B}, 125 (1980).

\bibitem{DS} S.~Dimopoulos and L.~Susskind, Nucl.\ Phys.\ {\bf B155}, 237
(1979).

\bibitem{DRK} S.~Dimopoulos, S.~Raby, and G.~Kane, Nucl.\ Phys.\ {\bf
B182}, 77 (1981).

\bibitem{DE} S.~Dimopoulos and J.~Ellis, Nucl.\ Phys.\ {\bf B182}, 505
(1981).

\bibitem{PT} M.~Peskin and T.~Takeuchi, Phys.\ Rev.\ Lett.\ {\bf65}, 964
(1990); B.~Holdom  and J.~Terning, Phys.\ Lett.\ {\bf247B}, 88 (1990);
M.~Golden and L.~Randall, Nucl.\ Phys.\ {\bf B361}, 3 (1991).

\bibitem{WT} B.~Holdom, Phys.\ Rev.\ D {\bf24}, 1441 (1981); Phys.\
Lett.\ {\bf150B}, 301 (1985); K.~Yamawaki, M.~Bando, and K.~Matumoto,
Phys.\ Rev.\ Lett.\ {\bf56}, 1335 (1986); T.~Appelquist, D.~Karabali,
and L.~Wijewardhana, Phys.\ Rev.\ Lett.\ {\bf57}, 957 (1986);
T.~Appelquist and L.~Wijewardhana, Phys.\ Rev.\ D {\bf36}, 568 (1987).

\bibitem{WAG} C.~Wagner, MPI-Ph/93-25 (1993), to appear in the proceedings
of {\it Properties of SUSY Particles}, Erice, 1992.

\bibitem{WW} S.~Weinberg, Phys.\ Rev.\ Lett.\ {\bf40}, 223 (1978);
F.~Wilczek, Phys.\ Rev.\ Lett.\ {\bf40}, 279 (1978).

\bibitem{D} S.~Dimopoulos, Nucl.\ Phys.\ {\bf B168}, 69 (1980);
M.~Peskin, Nucl.\ Phys.\ {\bf B175}, 197 (1980); J.~Preskill, Nucl.\ Phys.\
{\bf B177}, 21 (1981).

\bibitem{G} H.~Georgi, Nucl.\ Phys.\ {\bf B156}, 126 (1979).

\bibitem{RDS} S.~Raby, S.~Dimopoulos, and L.~Susskind, Nucl.\ Phys.\
{\bf B169}, 373 (1980).

\bibitem{EKK} E.~Eichten, K.~Kang, and I.-G.~Koh, J.\ Math Phys.\
{\bf23}, 2529 (1982).

\bibitem{DRS} S.~Dimopoulos, S.~Raby, and P.~Sikivie, Nucl.\ Phys.\ {\bf
B176}, 449 (1980).

\bibitem{BCS} P.~Bin\'etruy, A.~Chadha, and P.~Sikivie, Phys.\ Lett.\
{\bf 107B}, 425 (1981).

\bibitem{H} B.~Holdom, Phys.\ Rev.\ D {\bf 23}, 1637 (1981); D {\bf 24},
157 (1981); Phys.\ Rev.\ Lett.\ {\bf 57}, 2496 (1986);
Phys.\ Lett.\ {\bf 143B}, 227 (1984); {\bf 246B}, 169 (1990).

\bibitem{DR} S.~Dimopoulos and S.~Raby, Nucl.\ Phys.\ {\bf B192}, 353 (1981).

\bibitem{GEO} H.~Georgi, in {\it Architecture of Fundamental Interactions
at Short Distances}, Les Houches, 1985, eds. P.~Ramond and R.~Stora
(North-Holland, Amsterdam, 1987), p.~339.

\bibitem{AW} T.~Appelquist and L.~Wijewardhana, Phys.\ Rev.\ D {\bf 36},
568 (1987).

\bibitem{PS} J.~Pati and A.~Salam, Phys.\ Rev.\ D {\bf10}, 275 (1974).

\bibitem{E791} K.~Arisaka {\it et al}., Phys.\ Rev.\ Lett.\ {\bf70},
1049 (1993).

\bibitem{DJ} N.~Deshpande and R.~Johnson, Phys.\ Rev.\ D {\bf27}, 1193
(1983); J.~Ritchie and S.~Wojcicki, DOE-ER40757-016, CPP-93-16 (1993).

\bibitem{DRSW} M.~Dine, R.~Rohm, N.~Seiberg, and E.~Witten, Phys.\ Lett.\
{\bf 156B}, 55 (1985).

\bibitem{I} L.~Ib\'a\~nez, CERN-TH.5982/91, lectures presented at the 1990
CERN School of Physics, Mallorca, Spain, 1990;
CERN-TH.6342/91, talk given at the Workshop on Electroweak Physics
beyond the Standard Model, Valencia, Spain, 1991.

\bibitem{DALLAS}
S.~Weinberg, in {\it Proceedings of the XXVI International Conference on
High-Energy Physics}, Dallas, TX, 1992, ed. J. Sanford (American Institute
of Physics, New York, 1993), p.~346.

\bibitem{K} V.~Kaplunovsky, Nucl.\ Phys.\ {\bf B307}, 145 (1988); erratum
{\bf B382}, 436 (1992).

\bibitem{I2} L.~Ib\'a\~nez, Phys.\ Lett.\ {\bf126B}, 196 (1983);
L.~Ib\'a\~nez, D.~L\"ust, and G.~Ross, Phys.\ Lett.\ {\bf 272B}, 251 (1991).

\bibitem{AEKN}
I.~Antoniadis, J.~Ellis, S.~Kelley, and D.~Nanopoulos, Phys.\ Lett.\ {\bf
272B}, 31 (1991); S.~Kelley, J.~Lopez, and D.~Nanopoulos, Phys.\ Lett.\
{\bf 278B}, 140 (1992).

\bibitem{BL} D.~Bailin and A.~Love, Phys.\ Lett.\ {bf 278B}, 125 (1991);
{\bf 280B}, 26 (1992); {\bf 292B}, 315 (1992).

\bibitem{L} D.~Lewellen, Nucl.\ Phys.\ {\bf B337}, 61 (1990).

\bibitem{FIQ} A.~Font, L.~Ib\'a\~nez, and F.~Quevedo, Nucl.\ Phys.\
{\bf B345}, 389 (1990).

\bibitem{ANOM} A.~Font, L.~Ib\'a\~nez, and F.~Quevedo, Phys.\ Lett.\ {\bf
228B}, 79 (1989); C.~Geng and R.~Marshak, Phys.\ Rev.\ D {\bf 39}, 693 (1989);
D {\bf 41}, 717 (1990); J.~Minahan, P.~Ramond, and R.~Warner, Phys.\ Rev.\
D {\bf 41}, 715 (1990); K.~Babu and R.~Mohapatra, Phys.\ Rev.\ D {\bf 41}, 271
(1990); X.-G.~He, G.~Joshi, and R.~Volkas, Phys.\ Rev.\ D {\bf 41}, 278 (1990).

\end{thebibliography}
\end{document}